# Implementing RBAC model in An Operating System Kernel

Zhiyong Shan    SUN Yu-fang

(Institute of Software, The Chinese Academy of Sciences, Beijing 100080)

**This paper is in Chinese**

**Abstract**   In this paper, the implementation of an operating system oriented RBAC model is discussed. Firstly, on the basis of RBAC96 model, a new RBAC model named "OSR" is presented. Secondly, the OSR model is enforced in RFSOS kernel by the way of integrating GFAC method and Capability mechanism together. All parts of the OSR implementation are described in detail.

**Key words**    operating system，RBAC，access control

## 1  Introduction

  基于角色的访问控制（Role-Based Access Control 简称 RBAC）是九十年代兴起的一种有效的访问控制方法,是由美国国家标准化和技术委员会（NIST）的 Ferraiolo 等人在 1992 年提出的[1]。乔治麻省大学（George Mason University）的 Sandhu 教授于 1996 年又提出 RBAC96 模型[2]，成为经典的角色访问控制模型。

  操作系统安全是计算机安全系统的基础，"忽略操作系统安全的安全系统是建立在沙滩上的城堡"[4]。探讨如何在操作系统中实施 RBAC 是一项很有意义的工作，可以利用 RBAC 的以下优点[2][5][9]加强操作系统的安全性：

- 政策中立。在操作系统中支持多安全政策是操作系统安全的发展趋势[6][7]。以 RBAC 作为基础的访问控制机制，可以使操作系统支持多安全政策，而且，有利于不同操作系统之间的互操作[5]。
- 良好地支持最小特权安全原则(Principle of Least Privilege)[8]。使用户和应用程序只限于拥有完成其任务所必需的权限，防止滥用权限危害系统安全。
- 良好地支持职责分离（Separation of Duties）的安全原则。使相互冲突的两个职位之间不能相互兼任。如：限制系统管理员和安全管理员之间的兼任。
- 支持数据抽象。将对客体的操作许可抽象为更有意义的表达，而不仅仅是操作系统通常提供的读写执行等操作。
- 便于安全管理。使得安全管理员能够借助角色访问控制机制直观方便地描述和实施组织安全政策，同时也方便安全管理员进行权限管理。
- 但是，目前对 RBAC 的实现研究主要集中在应用层[2][3][23][24][25][26][27]，如：Web 服务器

和数据库。对于在操作系统中的实现研究比较少而且不完善[2][3]。

本文在实施 RBAC 之前,针对操作系统的特点对 RBAC96 模型进行细化和扩展。然后,采用 GFAC 和 Capability 机制相结合的办法在红旗安全操作系统内核中实现 OSR 模型。

## 2 扩展 RBAC96 模型

RBAC96 是一个模型族[2],包括 RBAC0、RBAC1、RBAC2 和 RBAC3 等模型,其中 RBAC3 是前三者的综合(见图 1)。RBAC3 包括以下内容:

1)用户(User)、角色(Role)、权限(Permission)、会话(Session)等四个实体。

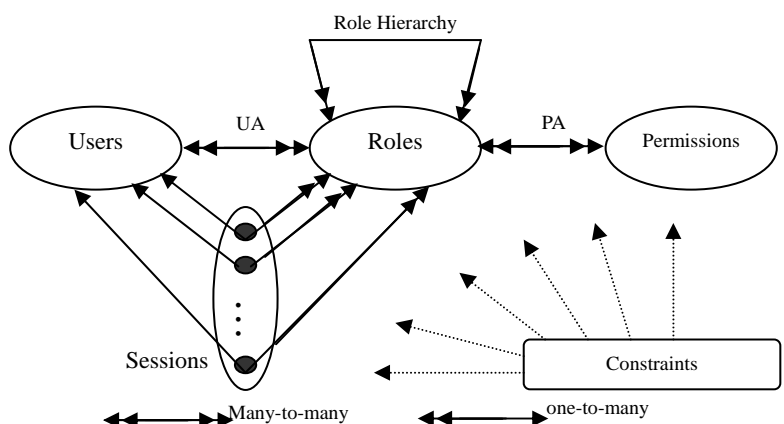

图 1 RBAC3 模型

2)权限分配(Permission Assignment)、用户分配(User Assignment)、角色继承(Role Hierarchy)和限制(Constraints)等四种关系。

3)用户→会话和会话→角色两种映射。

RBAC96 是一个高度抽象的通用模型,当应用到操作系统中时存在两点不足。

其一,可执行文件决定进程(进程对应于 RBAC96 的会话实体)的行为流程,应该参与权限的分配与传递,但是 RBAC96 模型并没有反映。

其二,操作系统中的权限比较复杂,RBAC96 模型缺乏细致的描述,不便于将模型实施。操作系统中所有的权限应该分为普通权限、应用层权限和特殊权限。

普通权限是主体访问客体前,在内核中实施检查的标志。通常的读、写文件等操作都是普通权限。

应用层权限是主体执行一项任务或操作前,在应用程序中实施检查的标志,这种任务或操作的粒度由应用程序的内在逻辑决定,可以是控制整个系统的操作,也可以是对某个文件一个字节的访问。

特殊权限是主体在执行一项维持安全操作系统正常而安全地运转的任务前,在内核中检查的标志,比如"改变(任何)文件的属主"、"网络管理"、"系统重起"等等权限都属于这种情况。可以分为系统管理特权、安全管理特权和审计特权。系统管理特权是维护系统正常运行的特权;安全管理特权是维护系统安全的特权;审计特权是进行系统安全审计的特权。这三种特权相互牵制相互监督。

Sandhu 教授在他的论文[22]中已经指出:RBAC 模型是一个开放式的概念模型,可以依据不同的应用进行扩展。所以,我们针对上述不足扩展 RBAC3,得到 OSR(Operating System oriented RBAC model)模型,见图 2。OSR 引入可执行文件作为实体,并且将操作系统所有权限分为普通权限、特殊权限和应用层权限。这样,角色可以包含三种权限,可以被分配给用户和可执行文件,进程同时从用户和可执行文件继承角色。

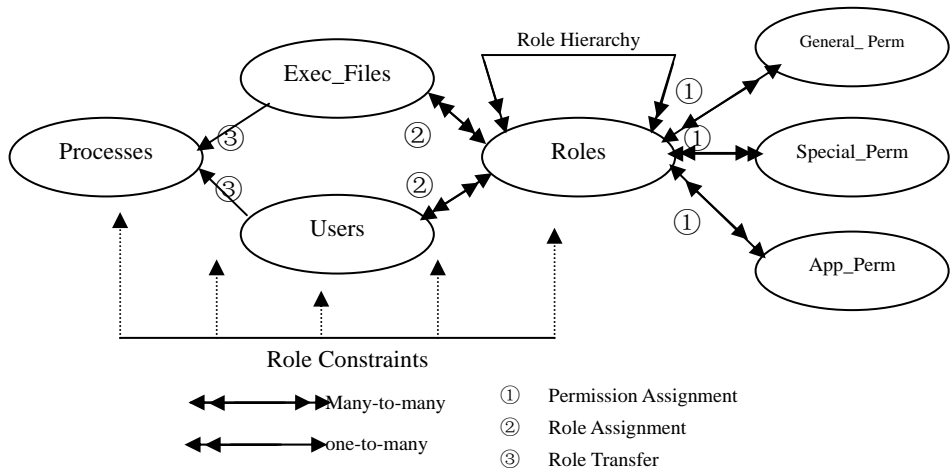

图 2 OSR 模型

## 2 OSR 模型实现

我们采用 GFAC 与 Capability 机制相结合的方法在 RFSOS 内核中实施 OSR 模型。通用访问控制框架 GFAC 是一种支持多安全政策模块的访问控制实施方法[10]，利用它可以方便地向可信计算机系统加入新的安全政策模块，而且也很容易验证新安全政策模块的正确性[11]。Capability 机制是 Linux 内核中的一种访问控制机制，将超级用户的特殊权限分割成一组特权，每个特权对应一个 Capability，进程依据其拥有的 Capability 决定它可以完成哪些特权操作。利用 Capability 机制可以实现对超级用户特权的控制和管理。

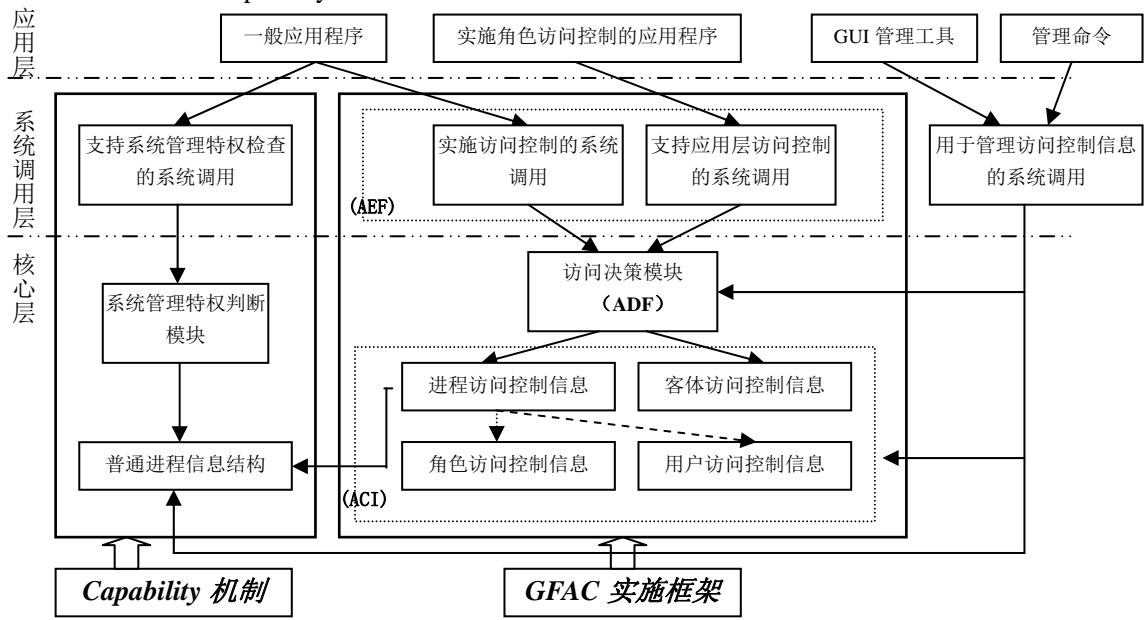

图 3 OSR 模型在 Linux 中的总体实施框架

总体的实施框架见图 3。主要分为 Capability 机制和 GFAC 实施框架两部分，Capability 机制部分完成对系统管理特权（即原来超级用户的特权）的访问控制，GFAC 部分完成对普通权限、应用层权限、安全管理特权和审计特权的访问控制。另外有安全管理部分，由 GUI 管理工具、管理命令和安全管理系统调用组成，主要用于管理访问控制信息。下面依据图 3 介绍 OSR 模型的实施。

## 3.1 GFAC 实施部分

依据GFAC方法，可信计算基（Trusted Computing Base TCB）由访问控制实施部分（Access Enforcement Facility AEF）和访问控制决策部分（Access Decision Facility ADF）组成。ADF实施了若干系统安全政策和一个元政策，决定进程的访问请求是否被许可。AEF 依据 ADF 的决策结果来控制安全相关系统调用的执行。

在 OSR 模型的实现中，内核的访问控制系统分为三个部分：AEF 部分、ADF 部分和 ACI（Access Control Information）部分。ACI 部分管理访问控制信息（ACI，即：主客体的安全属性）。图 4 表达了三个部分之间的关系：

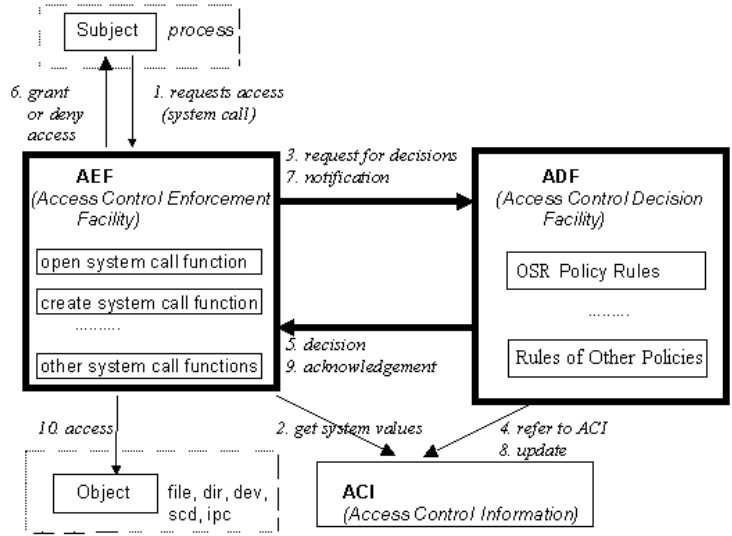

图 4　GFAC 框架的实施

发生访问时，AEF 从安全相关的系统调用中发送一个访问请求给 ADF。访问请求的参数包括：请求类型、发送请求进程的标识、要求访问客体的标识。ADF 先依据请求类型、政策规则和相关访问控制信息判断当前请求是否满足各个安全政策，然后由元政策综合各个安全政策的判断结论计算出最终的结果。若允许访问，则系统调用继续执行，并且在成功执行后由 AEF 通知 ADF 修改相应的访问控制信息。若禁止访问，则系统调用中断执行，并且返回一个错误给进程。

### 3.1.1 AEF 的实现

AEF 部分是通过在所有安全相关的系统调用中添加访问请求和修改通知来实现的。OSR 实施中，系统调用和访问请求的对应关系如表 1：

Table 1

表　1

| ADF request | System calls | ADF request | System calls |
|---|---|---|---|
| ADD_TO_KERNEL | Create_module(), init_module() | READ | Readdir(), readlink(), getdent() |
| ALTER | Ipc(), msgctl(), shmctl() | READ_ATTRIBUTE | Rslx_get_attr() 注：新增用于安全管理的系统调用 |
| APPEND_OPEN | Open(), msgsnd() | READ_OPEN | Ipc(), open(), msgrcv(), shmat() |
| CHANGE_GROUP | Chgrp(), fchgrp(), setgid(), setfsgid(), setregid(), setgroups() | READ_WRITE_OPEN | Ipc(), socketcall(), open(), shmat() |
| CHANGE_OWNER | Chown(), fchown(), setuid(), setfsuid(), setreuid() | REMOVE_FROM_KERNEL | Delete_module() |
| CHDIR | Chdir(), fchdir() | RENAME | Rename() |
| CLONE | Fork(), clone() | SEARACH | Kernel-internel |
| CREATE | Create(), ipc(), socketcall(), mkdir(), mknod(), symlink(), open(), msgget(), shmget() | SEND_SIGNAL | Kill() |

| DELETE | Ipc(), socketcall(), rmdir(), unlink(), msgctl() | SHUTDOWN | Reboot() |
| --- | --- | --- | --- |
| EXECUTE | Exec_ve() | SWITCH_LOG | Rslx_adf_log_switch() 注：新增用于开关 ADF 日志的系统调用 |
| GET_PERMISSIONS_DATA | Access() | SWITCH_MODULE | Rslx_switch() 注：新增用于开关安全政策模块的系统调用 |
| GET_STATUS_DATA | Stat(), fstat(), lstat(), new_stat(), new_fstat(), new_lstat(), statfs(), fstatfs(), msgctl(), shmctl() | TERMINATE | exit() |
| LINK_HARD | Link() | TRACE | Ptrace() |
| MODIFY_ACCESS_DATA | Utime() | TRUNCATE | Open(), truncate(), ftruncate() |
| MODIFY_ATTRIBUTE | Rslx_set_attr() 注：新增用于安全管理的系统调用 | UMOUNT | Umount() |
| MODIFY_PERMISSIONS_DATA | Chmod(), fchmod(), ioperm(), iopl() | WRITE | Rename() |
| MODIFY_SYSTEM_DATA | Adjtimes(), stime(), settimeofday(), sethostname(), setdomainname(), setrlimit(), swapon(), swapoff(), syslog() | WRITE_OPEN | Open() |
| MOUNT | Mount() | CHECK_APP_RIGHT | Rslx_rac_check_app_right() 注：新增用于检查应用层权限的系统调用 |

### 3.1.2 ACI 部分的实现

ACI 部分的作用是管理访问控制信息。访问控制信息主要包括主体、客体和角色的安全属性。访问控制信息的读取、添加、删除和修改都必需通过 ACI 函数来进行，任何其它方式都无法访问。与 OSR 模型实施相关的主要访问控制信息如表 2：

Table 2
表 2

| | 安全属性名 | 解释 | | 安全属性名 | 解释 |
| --- | --- | --- | --- | --- | --- |
| 角色 | child_roles | 直接子角色列表,表达角色继承关系 | 进程 | Rac_types | 进程所属类型列表 |
| | static_conflict__roles | 静态冲突角色列表 | | Max_roles | 进程最大角色集合 |
| | dynamic_conflict_roles | 动态冲突角色列表 | | Active_roles | 进程活动角色集合 |
| | Fd_right_vectors_array | 对文件目录客体的操作权限列表 | 用户 | Max_roles | 用户最大角色集合 |
| | Dev_right_vectors_array | 对设备文件客体的操作权限列表 | | Active_roles | 用户活动角色集合 |
| | proc_right_vectors_array | 对进程客体的操作权限列表 | 文件目录 | Rac_types | 文件目录所属类型列表 |
| | Ipc_right_vectors_array | 对进程间通信客体的操作权限列表 | | Exec_file_roles | 角色列表（只对可执行文件有效） |
| | scd_right_vectors_array | 对系统控制数据的操作权限列表 | 设备 | Rac_types | 设备所属类型列表 |
| | secadm_privileges | 安全管理特权列表 | | | |
| | sysadm_priviliges | 系统管理特权列表 | 进程间通信 | Rac_types | 进程间通信所属类型列表 |
| | audadm_priviliges | 审计特权列表 | | | |
| | app_privileges | 应用层权限列表列表 | | | |

由上表可以看出，角色的权限由九个权限列表所刻画。三个特殊权限和应用层权限列表都是权限向量，每一位代表一个权限。普通权限的五个权限列表都是权限向量组，组中的每个权限向量代表该角色对一种类型客体的权限集合，权限向量的每一位代表一个权限。

由于进程可以同时拥有很多角色，从效率的角度考虑，我们在实际实现中将进程所有激活角色的权限合并起来存放在进程访问控制信息中，以便 ADF 能够快速地判断进程是否有权限。

每种 SCD（系统控制数据）都有其固有的类型，所以不必保存 SCD 的类型信息。

在内存中，系统所有的角色以数组的形式存在，每个数组元素保存一个角色；文件、目录和设备均是以双向链表的形式存在，链表分两个层次组织，第一层双向链表是设备链表，第二层双向链表是客体链表（客体包括文件、目录和设备），客体链表的每个元素保存一个客体的 ACI，整个客体链表就保存着同一存储设备下所有客体的 ACI，客体链表的头指针放

在设备链表的对应元素中；进程、进程间通信和用户的 ACI 分别由一个单层的双向链表保存。

在磁盘上，除进程和进程间通信外，所有主体和客体的 ACI 都以文件的形式保存在一个特殊保护的目录下，无法直接访问该目录，只有安全管理员通过内核提供的专用系统调用才能访问。内存中发生改变的 ACI 定时地被刷到磁盘上去。

ACI 部分向外提供的接口主要有四个：从磁盘上读 ACI 链表，向磁盘写 ACI 链表，读指定主客体的指定安全属性的值，修改指定主客体的指定安全属性。

### 3.1.3 ADF 的实现

OSR 模型的 ADF 实施部分可以用表 3 描述。其中请求 R_MAC_XX 和 R_IAC_XX 是从 ADF 的保密性强制访问控制和完整性访问控制政策模块发送来的，R_AUDIT_XX 是从审计系统调用以及审计守护进程发送来的，R_APPLICATION 是从应用层权限检查的系统调用发送来的。

Table 3
表 3

| Request \ target | T_FILE | T_DIR | T_DEV | T_PROCESS | T_IPC | T_SCD |
|---|---|---|---|---|---|---|
| R_ADD_TO_KERNEL | CR | | | | | |
| R_ALTER | | | | | CR | |
| R_APPEND_OPEN | CR | | CR | | CR | |
| R_READ_WRITE_OPEN | CR | | CR | | CR | |
| R_CHANGE_GROUP | CR | CR | | | CR | |
| R_CHANGE_OWNER | CR | CR | | ①，SR, ST | CR | |
| R_CHDIR | | CR | | | | |
| R_READ | | CR | | | | |
| R_SEARACH | | CR | | | | |
| R_WRITE | | CR | | | | |
| R_CLONE | | | | ②，SR, ST | | |
| R_CREATE | | ③, ST | | | ④，ST | |
| R_DELETE | CR | CR | | | CR | |
| R_EXECUTE | CR | | | ⑤，SR, ST | | |
| R_GET_PERMISSIONS_DATA | | | | | | |
| R_GET_STATUS_DATA | | | | | | CR |
| R_LINK_HARD | CR | | | | | |
| R_TRUNCATE | CR | | | | | |
| R_MODIFY_ACCESS_DATA | CR | CR | | | | |
| R_RENAME | CR | CR | | | | |
| R_MODIFY_ATTRIBUTE | CP_sec | | | | | |
| R_MODIFY_PERMISSIONS_DATA | CR | CR | | | CR | CR |
| R_MODIFY_SYSTEM_DATA | | | | | | CR |
| R_MOUNT | | CR | CR | | | |
| R_READ_ATTRIBUTE | CP_sec | | | | | |
| R_READ_OPEN | CR | CR | CR | | CR | |
| R_REMOVE_FROM_KERNEL | | | | | | |
| R_SEND_SIGNAL | | | | CR | | |
| R_TRACE | | | | CR | | |
| R_SHUTDOWN | | | | | | |
| R_SWITCH_LOG | | | | | | |
| R_SWITCH_MODULE | | | | | | |
| R_TERMINATE | | | | CR | | |
| R_WRITE_OPEN | CR | | CR | | CR | |
| R_UMOUNT | | CR | CR | | | |
| R_AUDIT_STOP | CP_aud | | | | | |
| R_AUDIT_SAVE_CONFIG | CP_aud | | | | | |
| R_AUDIT_RELOAD_CONFIG | CP_aud | | | | | |
| R_AUDIT_WORK | CP_aud | | | | | |
| R_AUDIT_START | CP_aud | | | | | |

| | |
|---|---|
| R_MAC_ADD_TO_KERNEL | |
| R_MAC_MOUNT | |
| R_MAC_SHUTDOWN | CP_sys |
| R_MAC_REMOVE_FROM_KERNEL | |
| R_MAC_UMOUNT | |
| R_IAC_MODIFY_ATTRIBUTE | |
| R_IAC_READ_ATTRIBUTE | |
| R_MAC_GET_STATUS_DATA | |
| R_MAC_MODIFY_ATTRIBUTE | |
| R_MAC_MODIFY_PERMISSIONS_DATA | |
| R_MAC_READ_ATTRIBUTE | CP_sec |
| R_MAC_SWITCH_LOG | |
| R_MAC_SWITCH_MODULE | |
| R_APPLICATION | CP_app |

注：

1. 以 CR 表示：检查进程对指定客体是否有普通访问权限；以 CP_sec 表示：检查进程是否有安全管理特权；以 CP_sys 表示：检查进程是否有系统管理特权；以 CP_aud 表示：检查进程是否有审计特权；以 CP_app 表示：检查进程是否有应用层权限。

2. 以 SR 表示：访问结束后，重新设置进程的角色集合；以 ST 表示：访问结束后，重新设置进程或客体的类型列表；

3. ①表示：判断静态角色冲突，若两个用户间或新用户和可执行文件间存在静态角色冲突则拒绝；②表示：检查进程是否对新进程所在的类型有 CREATE 权限；③表示：先检查是否有权限在该目录下创建客体，然后检查用户是否有权创建用户安全属性所指定的缺省客体类型；④表示：检查是否有权创建由用户安全属性指定的 IPC 类型；⑤表示：判断可执行文件和用户之间是否存在静态角色冲突，若是则拒绝。

ADF 的决策流程如图 5：

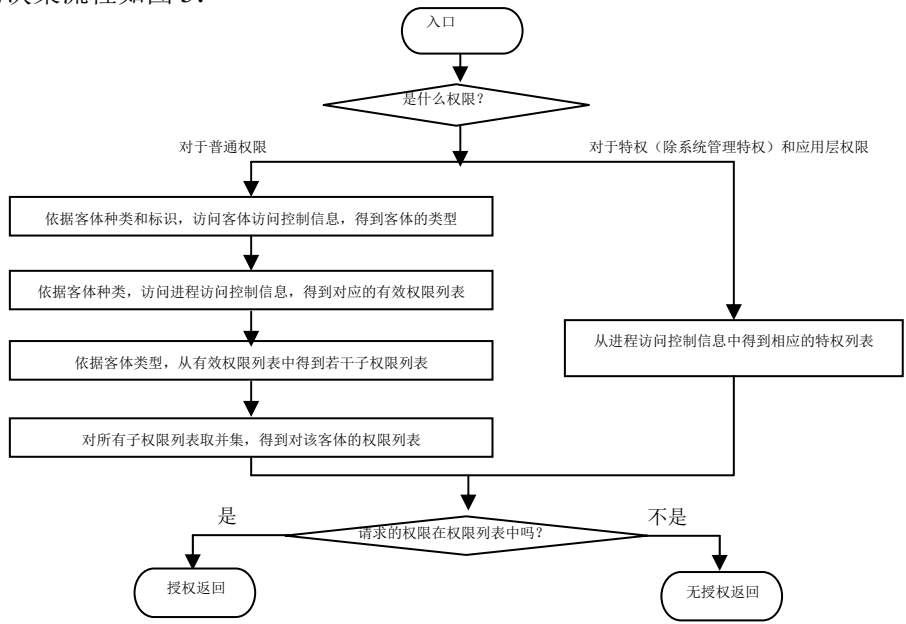

图 5  ADF 决策流程

### 3.2  Capability 实施部分

Linux 中的 Capability 机制是依据 POSIX 标准草案 POSIX draft 1003.1e[12]来实现的。Capabilities 是将超级用户的所有特殊权限分割后得到的一组权限。进程有三个 Capabilities 权限向量。在进程执行中，依据它的一个权限向量 effective capabilities 决定它当前拥有哪些超级用户特权。进程在产生时完全继承父进程的三个 Capabilities 向量，执行新映像文件后，

按照一定的规则改变进程的三个 Capabilities 向量。

我们在 OSR 模型的实现中，只利用进程的 effective capabilities 权限向量，并且去除 Capability 机制的 Capabilities 向量变化部分。每当进程的活动角色集合发生变化，都依据活动角色集合计算出进程的系统管理特权，并立即更新 effective capabilities 权限向量。需要检查进程的超级用户特权时，内核依据 effective capabilities 权限向量进行判断。这样，进程当前活动角色集合所确定的系统管理特权，成为 Capability 机制实际判断权限的依据，就将 Capability 机制和 OSR 模型的实现统一起来了。

### 3.3 系统缺省状态的确定

缺省状态的目的是保证系统能够顺利无碍运行的同时，对系统进行一些基本的保护。如下：
1．系统预设的客体类型及分配
- 缺省型：缺省状态下，除安全型和审计型外，所有其他客体都属于此类型
- 安全型：保存安全属性信息的文件和目录，安全管理工具和安全管理的命令
- 审计型：审计文件及目录，审计工具和命令

2．缺省的角色及角色关系
- 四个特权角色：可信系统管理员、系统管理员、安全管理员和安全审计员。它们的权限是不能被修改的。可信系统管理员拥有系统内所有的权限，但是不能授予任何用户，只能由内核授予系统进程
- 通用角色：拥有对缺省型客体的访问权限。设置这个角色的目的是使系统能运行起来
- 系统管理员、安全管理员和审计管理员之间存在静态冲突关系

3．新建主客体的类型
- 可以由用户在新建前指定
- 对于文件、目录、设备将继承父目录的类型
- 对于进程间通信将继承进程的类型
- 对于进程将由属主用户、父进程和可执行文件共同决定其类型

4．新建主客体的角色
- 新建用户的角色都是通用角色
- 新建进程的情况比较复杂：刚产生时，其角色集合从父进程继承，装入新的可执行文件时，其角色是可执行文件的角色集合与用户的活动角色集合（用户活动角色集合是指用户最大角色集合中当前生效的角色的集合）的并集
- 新建可执行文件的角色集合为空

### 3.4 继承关系和限制关系的实现

角色继承关系是 OSR 模型描述组织层次关系的关键，由角色的安全属性 child_roles 来刻画。继承关系有两点要求：
- 子角色继承父角色的所有权限；
- 不能出现循环继承。

如何确保满足了这两点要求是实现角色继承关系的关键。我们在两个位置进行检查：
- 设置角色的 child_roles 安全属性时，检查该角色的权限是否被每个子角色所包含，检查是否存在循环继承。
- 将角色分配给用户和进程时，检查得到的用户或进程的角色集合中，是否有存在继承关系的角色，排除父角色。

角色限制关系是 OSR 模型实现职责分离安全原则的关键，由角色的安全属性

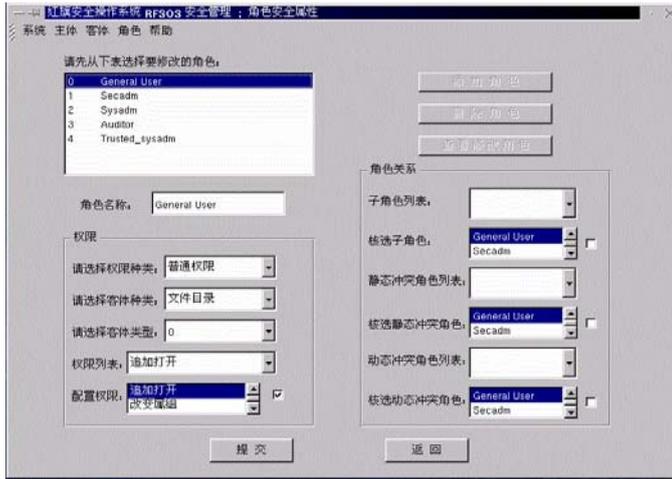 static_conflict_roles 和 dynamic_conflict_roles 刻画。在设置角色安全属性、分配角色、激活角色时都必须进行检查。

### 3.5 安全管理（系统调用、命令和图形工具）

对 OSR 模型的安全管理可以分为三个层次：系统调用层、管理命令层和图形工具。系统调用层属于内核的一部分，为管理命令和图形工具的开发提供支持。

图 6  安全管理图形工具

新加的系统调用主要有：
- rfsos_get_attr()：获取主客体的安全属性
- rfsos_set_attr()：设置主客体的安全属性
- rfsos_osr_add_del_role()：增加或删除角色
- rfsos_osr_get_role_attr()：获取角色的安全属性
- rfsos_osr_set_role_attr()：设置角色的安全属性
- rfsos_osr_activate_role()：激活角色
- rfsos_osr_check_app_right()：检查主体的应用层权限

新加的安全管理命令主要有：
- rfsos_get_user_attr, rfsos_get_proc_attr, rfsos_get_file_dir_attr, rfsos_get_ipc_attr, rfsos_get_dev_attr：获取指定主客体的安全属性
- rfsos_set_user_attr, rfsos_set_proc_attr, rfsos_set_file_dir_attr, rfsos_set_ipc_attr, rfsos_set_dev_attr：设置主客体的安全属性
- rfsos_osr_add_role, rfsos_osr_del_role：增加和删除角色
- rfsos_osr_get_role_attr, rfsos_osr_set_role_attr：获取和设置角色安全属性
- rfsos_osr_activate_role：激活角色

图 6 是安全管理图形工具中，角色安全属性管理的窗口。

## 4  相关工作及比较

在操作系统中实施 RBAC 的研究比较有影响的系统有 Trusted Solaris 8、SE-Linux、LOCK6 和其它一些实验系统。

Trusted Solaris 8[13][14]中借助用户帐号来实现角色，是商业化比较成功的例子，实现方式简单，但是有其不可克服的缺点。首先，角色之间无法实现继承关系，而继承是 RBAC 模型的重要特点。其次，在同一时刻进程只能激活一个角色，激活的角色用户将代替进程原来的属主，所以，用户在系统中工作可能要反复地在角色和用户之间、角色和角色之间切换，而每一次切换都需要口令认证，显然给用户增加了操作负担。如果要减少切换次数，势必增加角色的权限（将多个角色的权限合并），这就违反了最小特权原则。由论文[2]也可以知道，允许进程同时激活多个角色是 RBAC 模型支持最小特权原则的关键。另外，Trusted Solaris 8 中的 RBAC 只是实现了对超级用户特权的访问控制，没有实现对操作系统所有权限的控制，

所以其作用是有限的。

LOCK6[3]借助 Type Enforcement 机制实现 RBAC，但是没有真正在内核底层支持角色访问控制机制。以 Type Enforcement 访问控制机制为基础实现 RBAC，实际就在角色和权限之间增加了一层：访问控制域（domain），使系统变得更复杂了。因为，LOCK6 中，一个进程同时具有属主用户、角色和访问控制域三个可以决定其权限的属性，必须维护三者之间一致性，也就是说进程的这三个属性必须同时与系统的安全属性数据库一致：当前的角色是否在用户允许的角色集合中，当前的访问控制域又是否在角色允许的域集合中。其次，没有在内核中支持继承和职责分离等要求。再其次，在同一时刻进程只能激活一个角色，也就存在着与 Trusted Solaris 8 类似的问题。最后，LOCK6 不是一个主流的操作系统，所以限制了它的实用意义。

SE-Linux[15]是支持安全政策多样性和动态性最著名的安全操作系统，它在 FLASK 结构下实现 RBAC，但只支持"角色迁移"和"继承"，没有提供直接的机制支持角色间的限制关系，如角色间的互斥关系，所以它不便表达职责分离原则。另外，在同一时刻进程也只能激活一个角色，存在着与 Trusted Solaris 8 类似的问题。

本文的研究相比之下有几个优点：

- 直接在操作系统内核中实施 RBAC，而不是在其它机制的基础上实施（如 Trusted Solaris 8 借助用户帐号表达角色、LOCK6 借助 Type Enforcement 机制支持 RBAC），使系统的访问控制更简洁，而且便于以角色访问控制机制为基础支持多安全政策[16]。
- 对经典的 RBAC 模型做了面向操作系统的扩展。角色赋给可执行文件，可以更好地限制应用程序的权限范围，防止恶意盗用和滥用权限，有利于贯彻最小特权原则；能够控制整个操作系统的所有权限，包括普通权限、系统管理特权、安全管理特权、审计特权和应用层权限，符合访问监控机"监控所有访问"的要求[18]，也更有利于实现最小特权原则。
- 所实现的 RBAC 特性更全面，能更好的支持基于角色的访问控制安全政策。依据美国国家标准化和技术委员会（NIST）2000 年 12 月提出的 RBAC 标准草案[15]，RBAC 实施中应该支持的特性包括：基本 RBAC 要求、继承关系、静态职责分离关系和动态职责分离关系。在本文的工作中已经基本实现。
- 利用 GFAC 和 Capability 相结合的方法实现 RBAC。在对所有权限进行控制的前提下，最大限度地减少了对内核的修改，特别是对内核关键数据结构的修改。因为，ADF 和 ACI 是独立添加的部分，AEF 实际也只是在系统调用中添加了 ADF 访问请求和 ADF 属性设置通知。经过实际测试，在缺省配置下（实现了系统管理员、安全管理员、审计管理员和普通用户角色），所有应用程序都能顺利运行。同时，这种实现方法较好地结合了 GFAC 和 Capability 机制，将超级用户的特权纳入角色机制进行统一控制，从而真正达到对 RFSOS 操作系统所有权限的控制。

本文的研究和实施工作已经加入红旗安全操作系统*3.0 版。希望该文能够对角色访问控制模型和安全操作系统的研究有所裨益。欢迎阅读本人的其他论文[29-67]。

---

\* 红旗安全操作系统是由中国科学院软件研究所和中科红旗软件公司联合开发的基于 Linux 的安全操作系统[28]，于 2001 年 6 月通过公安部组织的国家信息安全评价标准第三级认证，于 2002 年 1 月通过中国科学院组织的"基于国际/国家标准的安全操作系统"（中科院成鉴字[2002]第 003 号）鉴定会的鉴定。